\documentclass[journal,twoside]{IEEEtran}

\usepackage[normalem]{ulem}
\usepackage[table,xcdraw]{xcolor}
\ifCLASSINFOpdf
  \usepackage[pdftex]{graphicx}
  \graphicspath{{./figs-blind/}}
  
  \DeclareGraphicsExtensions{.png}
\else
  \usepackage[dvips]{graphicx}
  \graphicspath{{./figs/}}
  \DeclareGraphicsExtensions{.eps,.mps}
\fi
\usepackage[cmex10]{amsmath}

\ifCLASSOPTIONcompsoc
 \usepackage[caption=false,font=normalsize,labelfont=sf,textfont=sf]{subfig}
\else
 \usepackage[caption=false,font=footnotesize]{subfig}
\fi
\usepackage{dblfloatfix}

\usepackage{url}
\usepackage{booktabs}
\usepackage{multirow}

\usepackage{textcomp}
\usepackage{enumitem}
\usepackage{xcolor}
\usepackage{comment}

\newcommand{\rem}[1]{\textcolor{red}{#1}}

\usepackage{float} % Only For Revision. Allow fix the position of a figure wit [H] option
\usepackage[abs]{overpic} % Superimpose cosas

\usepackage{psfrag} % Fonts substitution in pictures.

\usepackage{multicol}

\newcommand{\fig}[1]{Fig.~\ref{#1}}
\newcommand{\equ}[1]{Eq.~(\ref{#1})}
\newcommand{\tab}[1]{Table\,\ref{#1}}

\newcommand{\quotes}[1]{``{#1}''}

\newcommand{\ph}[1]{\phantom{#1}}

\newcommand{\ME}{\mathcal{E}} % Energy
\newcommand{\MP}{\mathcal{P}} % Power
\newcommand{\hide}[1]{{\iffalse {#1} \fi}}

\begin{document}
\title{Exploring the Performance Boundaries of NB-IoT}

\author{
    Borja~Martinez,~\IEEEmembership{Senior~Member,~IEEE,}
	Ferran~Adelantado,~\IEEEmembership{Member,~IEEE,}
	Andrea~Bartoli,
    Xavier~Vilajosana,~\IEEEmembership{Senior~Member,~IEEE}%
    \thanks{B.\,Martinez and F. Adelantado are with IN3 at Universitat Oberta de Catalunya.}% <-this % stops a space
	\thanks{A.\,Bartoli is with Worldsensing S.L.}%
    \thanks{X.\,Vilajosana is with IN3 at Universitat Oberta de Catalunya and Worldsensing S.L.}%
}

%\thanks{Copyright (c) 2012 IEEE. Personal use of this material is permitted. 
%However, permission to use this material for any other purposes must be obtained from the IEEE by sending a request to pubs-permissions@ieee.org}

\ifCLASSOPTIONpeerreview
  \markboth{IEEE Internet of Things Journal}%
  {Exploring the Performance Boundaries of NB-IoT}
\else
  %\markboth{IEEE Internet of Things Journal,~Vol.~XX, No.ZZ, Month~2018}{B.~Martinez \MakeLowercase{\textit{et al.}}: }

  \markboth{}%
  {Martinez \MakeLowercase{\textit{et al.}}: Exploring the Performance Boundaries of NB-IoT}

  %\markright{Martinez \MakeLowercase{\textit{et al.}}: Exploring the Performance Boundaries of NB-IoT}

\fi

\maketitle

% As a general rule, do not put math, special symbols or citations in the abstract or keywords.
\begin{abstract}
NarrowBand-IoT has just joined the LPWAN community. Unlike most of its competitors, NB-IoT did not emerge from a blank slate. Indeed, it is closely linked to LTE, from which it inherits many of the features that undoubtedly determine its behavior. In this paper, we empirically explore the boundaries of this technology, analyzing from a user's point of view critical characteristics such as energy consumption, reliability and delays. The results show that its performance in terms of energy is comparable and even outperforms, in some cases, an LPWAN reference technology like LoRa, with the added benefit of guaranteeing delivery. However, the high variability observed in both energy expenditure and network delays call into question its suitability for some applications, especially those subject to service-level agreements.
\end{abstract}


\IEEEpeerreviewmaketitle

%-----------------------------------------------------------------------
% INTRODUCTION
%-----------------------------------------------------------------------
% \input{nbiot-sec-intro}
% ======================================================================
\section{Introduction}
\label{sec:introduction}

\IEEEPARstart{A}{} highly diversified Internet of Things (IoT) ecosystem has been created as a result of the IoT hype and the corresponding increase in venture capital being injected into companies and start-ups. Multiple wireless technologies have been developed, some of them standardized (e.g., Bluetooth, IETF 6TiSCH, LoRaWAN, Weightless, and Sigfox, among others), and many proprietary alternatives are constantly being offered. All of them meet the requirements - or at least try to - established by some set of applications whose boundaries are shaped by technological constraints. Matching emerging applications with existing technologies has become one of the main challenges for IoT initiatives, especially when a new technology appears in the landscape and the map must be redrawn.

One of the first IoT applications that showed a clear value proposition was smart metering. Non-intrusive remote access to utility meters delivered the ability to reduce the intervals between readings, thus enabling new services for users (such as dynamic pricing and usage patterns analysis) and operators (such as load balancing between multiple users). The almost immediate success is partially responsible for the preconception that IoT applications should be low power and low data rate. This latent bias persists today.

\begin{figure}[ht!]
  \centering
  \includegraphics[width=1.0\columnwidth]{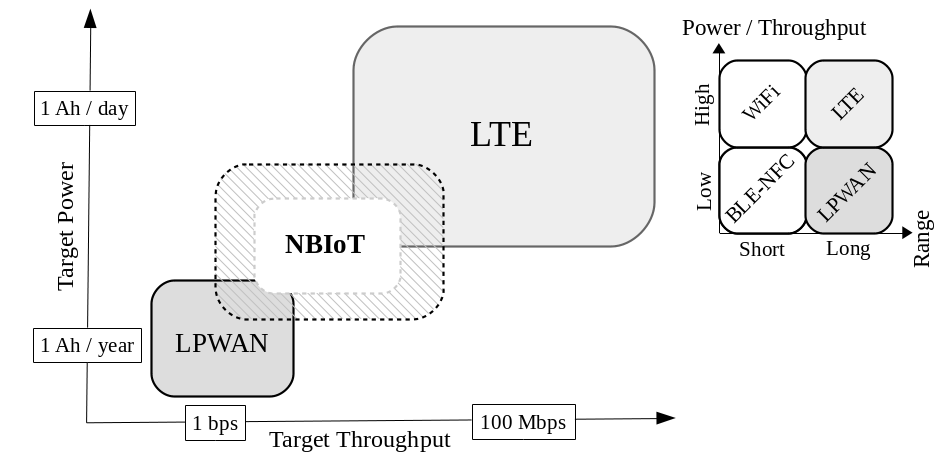}
  \caption{NB-IoT positioning. NB-IoT is 3GPP proposal for addressing the emerging long-range, low-power, low-data-rate IoT market.}
  \label{fig:e1}
\end{figure}

As an extension of Long Term Evolution (LTE), Narrow Band IoT (NB-IoT) was conceived within that framework and reflects a set of specifications particularly well suited to the smart metering use case. The 3GPP standards body focused on enhancing the characteristics of the User Equipment (UE) \cite{3gpp.45.820} in order to face the new IoT market (\fig{fig:e1}).  
This resulted in the definition of two new device categories, namely Cat NB1 and Cat NB2, both of which are characterized by limited radio transmission/reception and radio access capabilities \cite{3gpp.36.101,3gpp.36.306}.

For example, compared with LTE, some constraints were relaxed: NB-IoT devices are seen as stationary, only small chunks of data are intermittently transmitted and applications are envisaged as delay-tolerant; while other features were reinforced: a huge number of devices to accommodate (several orders of magnitude beyond LTE devices), often installed at places with poor coverage (e.g., the basements of buildings) and/or without power supply (which basically implies battery-operated UEs with reasonable lifetimes \cite{3gpp.36.300}).

In this article, we explore the boundaries that resulted from this approach and place special emphasis on the drawbacks attributable to the LTE legacy while also discussing optimizations that specifically target the IoT. To this end, we take a pragmatic perspective by taking the position of a potential adopter of the technology and focusing on those parameters that fall within end-user control. Under this premise, we:

\begin{enumerate}
  \item Analyze the main characteristics at the core of NB-IoT, especially those oriented towards improving coverage and reducing power consumption.
  \item Conduct a thorough experimental characterization to reveal the behavior of NB-IoT devices in actual operation. 
  \item Set realistic boundaries for the technology based on the obtained results. In view of these limits, we question its suitability for different IoT applications and use cases.
  \item Compare NB-IoT to LoRaWAN, which we consider the most prominent technology at this moment within the Low Power Wide Area Network (LPWAN) ecosystem (see \fig{fig:e1}).
\end{enumerate}

% what is already done and why another model?
Our work serves to complement the existing state of the art. Several studies provide theoretical models not only for the energy consumption of NB-IoT networks \cite{Maldonado2018}, 
but also for their latency and delay bounds \cite{Feltrin18Survey}, 
impact of coverage extensions \cite{Maldonado18CES}, 
(theoretically) optimal configuration strategies \cite{Feltrin18NbIoTPerformance} 
and overall performance for particular verticals \cite{Soussi17performance}. 
However, none of these efforts focus on the adopter and present an operational and empirical analysis of the technology when it is deployed in a real network. We argue that, despite the unquestionable value of the theoretical models (for example, to understand orders of magnitude or to guess the theoretical upper and lower bounds), an empirical approach provides real insights into the variability that a UE experiences when deployed in real conditions. Our work therefore goes in this direction as a complement to related works, and it further provides empirical measurements for UEs that are deployed using a real-world NB-IoT network – always while taking the perspective of an application developer.

%article organization
The article is organized as follows.
Section~\ref{sec:nbiot} describes the power saving and coverage extension mechanisms designed in the LTE Release 13 for supporting IoT scenarios. 
Section~\ref{sec:behaviour} looks at the behavior of UE from an energy consumption point of view and under different realistic configurations.
Section~\ref{sec:performance} then presents the probabilistic energy consumption and latency distributions based on the empirical data gathered.
%obtained from the operation of UEs when connected to the Vodafone NB-IoT network.
Section~\ref{sec:comparative} compares the obtained results with those from LoRaWAN, a well-established LPWAN technology, in order to rank the NB-IoT technology from a practical perspective.
Finally, Section~\ref{sec:conclusion} concludes this article.

%-----------------------------------------------------------------------
% 2.- NB-IoT Optimizations
%-----------------------------------------------------------------------
% \input{nbiot-sec-optimizations}
% ======================================================================

\section{LTE optimizations for NB-IoT}
\label{sec:nbiot}

\begin{figure*}[t]
  \centering
  \includegraphics{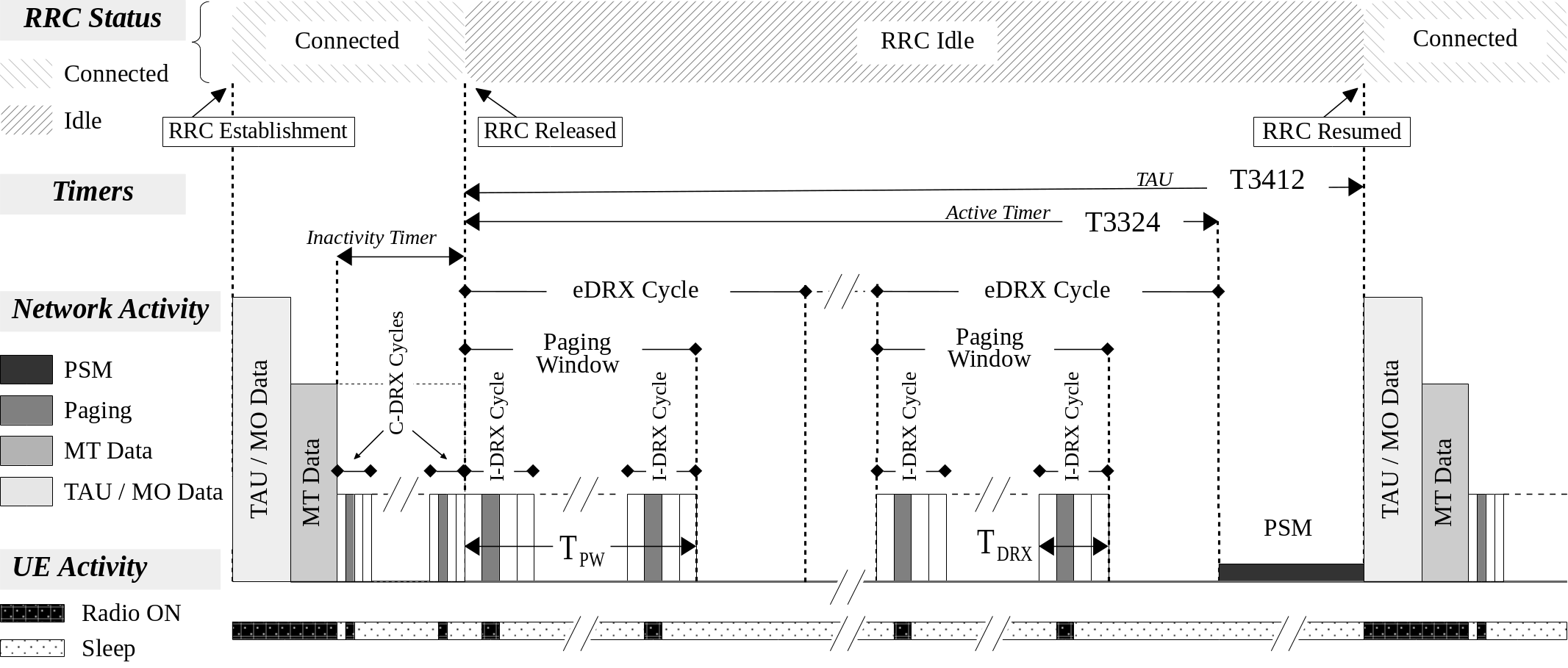}
  \caption{Summary diagram of UE's behavior in NB-IoT. 
  From top to bottom: 
  (top) state of the RRC connection,
  (middle) timing involved and 
  (bottom) radio interactivity between the UE and the network, 
  with the associated power consumption depicted schematically.}
  \label{fig:rrc-modes}
\end{figure*}

NB-IoT is well described in the literature \cite{Ratasuk2016,RohdeWhitePaper}. 
Hence, we will focus only on those LTE enhancements that are fundamental for understanding the NB-IoT operational trade-offs, especially those regarding  the low power and low data rates described in the previous section.

\subsection{Overview}
In order to illustrate the operations of NB-IoT and different power saving optimizations, we rely on \fig{fig:rrc-modes}. 
The LTE Radio Resource Control (RRC) protocol has only two states: RRC Connected and RRC Idle. In NB-IoT Release 13, the cell handover and redirection is not supported; so the state model of the RRC becomes quite simple. The figure shows (at the top) these two possible states. When the User Equipment (UE) is woken up for the first time, the network connection is established and the UE enters in the RRC Connected state. While connected, the UE can access the network and request communication resources.

At this point, LTE Release 13 has extended the configurability of the LTE power saving modes in order to support wider trade-offs in terms of energy consumption and the UE's communication capabilities. The power saving options in NB-IoT include several mechanisms:

\subsubsection{Connection Release/Resume}
When the base station, referred as Evolved Node B (eNB), releases the connection, the UE transits to the RRC Idle state and stores the current Access Stratum (AS) security context. The UE may later resume the RRC Connected state with that context, thus avoiding the AS setup and saving considerable signaling overhead for the transmission of infrequent small data packets. The main benefit of this mechanism is that it avoids renegotiating security, which is discussed in Section~\ref{subsec:security}.

\subsubsection{Idle-state power optimizations}

When the connection is released, the UE may enter one of the two saving modes: the Extended/Enhanced Discontinuous Reception Mode (eDRX) or the Power Saving Mode (PSM). Both modes are complementary and their goal is to reduce the overall energy consumption of the UE in the absence of traffic. The eDRX is designed to reduce the energy consumption of the UE while idle-waiting for downlink messages. In this mode, the UE does not have resources assigned but instead continues listening to  broadcast information by the network. This mechanism allows the UE to notice if there is any data available to be received, which would trigger an RRC Resume. When the eDRX expires, or it is forced to, the UE moves to PSM. In PSM, the device turns the radio off and is therefore unreachable by the network. However, this mechanism facilitates the device to enter deeper hardware sleep modes.

The remainder of this section delves into the details of these mechanisms, which are presented in a more technical language using LTE nomenclature. From a user/adopter perspective, this discussion can be passed over without losing any key insights into the results, which we begin discussing in Section \ref{sec:behaviour}.

\subsection{Power Saving Mechanisms in NB-IoT}

\subsubsection*{eDRX in Idle state}
%\textbf{eDRX: }

The DRX procedure is designed to efficiently support downlink communications. DRX can be (optionally) executed while the UE connection is in RRC Idle. In RRC Idle, new resources cannot be requested from the network, but the Narrowband Physical Downlink Control Channel (NPDCCH) is tracked to maintain network synchronization and to determine if there is downlink data pending. Energy efficiency arises from the paging mechanism: the UE monitors only some of the subframes, the Paging Occasions (PO) within a subset of radio-frames, and the Paging Frames (PF) \cite{3gpp.36.304}.
Paging therefore involves cycles that alternate between periods of active listening and sleep. Of course, this discontinuous reception incurs some additional latency, which is the price of saving energy.
In RRC Idle, DRX cycles of 128, 256, 512 and 1024 radio-frames are supported \cite{3gpp.36.331},
%[ETSI TS-136.331; defaultPagingCycle-r13].
ranging from 1.28\,s to 10.24\,s (each radio-frame spans 10\,ms).

The concept of extended/enhanced DRX (eDRX) in LTE is also applied in NB-IoT. If eDRX is supported, the time interval in which the UE does not monitor the paging messages may be considerably extended, by up to almost 3 hours. 
eDRX cycles have specific periods that are a multiple of the duration of a hyper-frame (1024 radio-frames, i.e., 10.24s). The eDRX process is controlled by a set of timers as defined in Table \tab{tab:timers}. 
In particular, the Active Timer (T3324) controls the time lapse during which the UE is reachable by the network in RRC Idle, i.e., the number of eDRX cycles. 
An eDRX cycle is composed of an active phase, controlled by a Paging Time Window (PTW) timer, which ranges from 2.56\,s to 40.96\,s followed by a sleep phase until the end of the eDRX cycle. Within the PTW, the standard LTE paging is observed.

\subsubsection*{eDRX in Connected state}

The DRX mechanism is not exclusive to RRC Idle. In RRC Connected, when there is no traffic, the UE also alternates active listening for POs and sleep periods. In Connected mode, DRX values are defined in multiples of subframes (1\,ms),
ranging from 10 to 2560 in LTE. 
In NB-IoT, the permitted values begin at 256 but are extended up to 9216, which is called “enhanced DRX” in RRC Connected mode (C-eDRX) \cite{3gpp.36.331}. 
%[ETSI TS-136.331, drx-Cycle-r13].
While still connected, the UE can access the network and request communication resources through the connectionless NB-IoT Physical Random Access Channel (NPRACH).

\subsubsection*{Power Saving Mode}

When the Activity Timer (T3324) expires (or the connection is released for other reasons), the UE enters in PSM mode. The PSM mode disconnects the radio completely, so the UE can enter a deeper sleep. While in PSM, the UE can resume the connection at any time. For that, it needs to initiate the resume process until it reaches the RRC Connected state. However, as we have already mentioned, the UE saves the context; therefore this process involves much less overhead than establishing a new connection. Obviously, as the radio is off, notifications will not be received by the UE during PSM. Therefore, the existence of downlink data will be noticed only when the connection is released. 

A timer, referred to as TAU or Extended Timer (T3412), is configured so that the UE wakes up periodically to perform a Tracking Area Update (TAU). The TAU process is analogous to that of LTE; however, it can be configured with a longer period of up to 413 days for NB-IoT \cite{3gpp.24.008}.% \todo{[ETSI TS-124.008, Table 10.5.163a (Timer 3)]}.

\begin{table*}[ht]
\centering
\caption{Summary of the eDRX timers.}
\label{tab:timers}
\resizebox{1.0\textwidth}{!}
{
\begin{tabular}{p{3.5cm}p{12cm}c}
\rowcolor[HTML]{C0C0C0} 
\multicolumn{1}{c}{\cellcolor[HTML]{C0C0C0}\textbf{Timer}} & \multicolumn{1}{c}{\cellcolor[HTML]{C0C0C0}\textbf{Description}} & \textbf{\begin{tabular}[c]{@{}c@{}}Configurable by UE?\end{tabular}} \\
\textbf{Inactivity Timer} & When the inactivity timer expires, it causes a transition from RRC Connected to the RRC Idle state. This timer is controlled by the eNB, not by the UE. & No \\
\textbf{Active Timer (T3324)} & The T3324 determines the duration during which the device remains reachable for the downlink through eDRX (RRC Idle mode). The device starts the Active Timer when it moves from RRC Connected to RRC Idle mode. When the Active Timer expires, the device moves to Power Saving Mode (PSM). & Yes \\
\textbf{Paging Time Window (PTW)} & This is the duration of a paging event composed of multiple DRX cycles. The paging event fits in the PTW; so the longer the PTW, the greater the number of DRX cycles. & Yes \\
\textbf{DRX Cycle} & The duration of a DRX cycle. It is a multiple of the Paging Occasions (PO) cycle (1280\,ms). In a DRX Cycle, the UE listens for one PO and sleeps for the following POs & No \\
\textbf{eDRX Cycle} & The duration of an eDRX cycle. (the time between two PTWs). & Yes
\end{tabular}%
} % ResizeBox
\end{table*}

%-------------------------------------------------------------------------------
%-------------------------------------------------------------------------------

\subsection{Coverage Enhancements}

NB-IoT is designed to support IoT devices that operate in deep indoor or remote areas \cite{3gpp.45.820}.
To satisfy these requirements the Release 13 enhancement introduces a set of techniques for improving coverage by taking advantage of the relaxed IoT requirements regarding data rate and latency. The improvement is estimated as a +20dB gain when compared to General Packet Radio Service (GPRS), which corresponds to a Maximum Coupling Loss (MCL) of 164dB \cite{Xu2018MCL}. 

To achieve this gain, two main mechanisms are introduced in \cite{3gpp.36.211}: %\todo{3GPPTS36.211}: 
repetitions\footnote{This is a concept different from that of retransmissions, 
which occur when it is noticed that a message has been lost.} 
and the ability to allocate variable bandwidth through the use of multi-tone operation. 

Repetitions occur in both uplink channels (e.g. NPRACH) and downlink channels (e.g. NPDCCH)
%Repetitions occur in the RACH \todo{to check that repetitions only occur in the RACH.} 
and they are determined by the eNB (the base station that connects directly to the UE) in accordance with the signal strength received from and reported by the UE. Based on that, the eNB establishes a category for the device, called Coverage Enhancement Level (ECL), which basically determines the number of repetitions (the number of repetitions in the uplink is limited to $2^i$, with $i=1 \ldots 7$).
There may be up to 3 levels, from ECL0 for normal operation to ECL2 for the worst case scenario. The network determines how CE levels are defined.

Multi-tone operation spans communication on multiple simultaneous subcarriers (1, 3, 6 and 12) while the mandatory single-tone communication uses a single subcarrier that extends transmissions over time.  
%
\begin{comment}
{In addition, the eNB determines dynamically the transmission power for the UE according to last received SNR and resource unit (RU) scheduling \cite{Maldonado18CES}.}
\end{comment}
{NB-IoT supports open-loop power control. The UE determines the transmission power according to the received Signal-to-Noise Ratio (SNR) and the transmission configuration done by the eNB \cite{Maldonado18CES,3gpp.36.213}.}

All this information is sent to the UE through a NPDCCH Downlink Control Information (DCI) object.
\begin{comment}
\rem{In the DCI, the start time of the Uplink Shared Channel (NPUSCH), the number of repetitions, the number of Resource Units (RUs) used for one transport block, the number of subcarriers and their position in the frequency band are informed. 
In addition the Modulation and Coding Scheme (MCS) index is transported in the DCI, providing extended information about the number of RUs,  the modulation scheme for the subcarrier RUs and the transport block size \cite{3gpp.36.213,RohdeWhitePaper}.}
\end{comment}
{In the DCI, the start time of the Uplink Shared Channel (NPUSCH), the number of repetitions, the resource assignment, 
the subcarrier indication and the Modulation and Coding Scheme (MCS) are informed \cite{3gpp.36.212}.}

\subsection{Security}
\label{subsec:security}

NB-IoT inherits the two security levels defined in LTE, namely Access Stratum (AS) and Non-Access Stratum (NAS) security. AS security is established between the UE and the eNB, whereas NAS security is established and managed by the UE and the Mobility Management Entity (MME). AS and NAS security are performed through a set of possible ciphering and integrity algorithms listed in \cite{3gpp.33.401}.
Specifically, 3GPP defines four ciphering algorithms, denoted as EPS Encryption Algorithms (EEA), and four integrity algorithms, known as EPS Integrity Algorithm (EIA)\footnote{The reader is referred to Annex B in \cite{3gpp.33.401} for further details on the ciphering and integrity algorithms}.
The configuration of each security level is performed after a negotiation between the UE and, respectively, the eNB or MME. Thus, the eNB and the MME have a prioritized list of ciphering and integrity algorithms, which are finally selected based on the security capabilities of the UE.
As for the configuration of security modes, NAS security is always configured before AS security.
Whereas both the encryption and integrity algorithms are applied in NAS, the ciphering in AS is used for RRC and User Plane (UP)while the integrity algorithm is applied only to RRC.% Separate configuration procedures/commands are required for each security level.
%Thus, in the access procedure, NAS security is always configured before AS security. } 
%In \cite{3gpp.33.401}, 3GPP defines the ciphering and integrity requirements for AS and NAS, known as }

An additional feature is that LTE defines Control Plane (CP) data transmission (inside RRC/NAS messages) as a lower overhead alternative to full Data Radio Bearer (DRB) IP user plane (UP) data transmission. For this data transfer method, security at the AS level is not applied and the RRC connection is not reconfigured.  
In this way the overhead is reduced considerably,
which makes it more suitable for short data transactions -
although it comes at the expense of diminishing transmission security, since it uses only the NAS security level.
This feature is mandatory in NB-IoT. 
{However, at the time this work was conducted, none of the evaluated platforms offered control over this functionality to the end user.}

%-----------------------------------------------------------------------
% 3.- Behavior
%-----------------------------------------------------------------------
%\input{nbiot-sec-behaviour}
% ======================================================================

\section{Observation of the UE Behaviour}
\label{sec:behaviour}
\captionsetup[subfigure]{width=0.48\textwidth}

\begin{figure*}[!t]
  \centering
    \subfloat[][The UE is listening for the control channel (NPDCCH) in C-DRX. 
    The network connection is released 20\,s after sending the datagram, 
    %when the inactivity timer expires.
    which is the expiration time of the Inactivity Timer.
    I-DRX is disabled.]{%
        \centering
        \label{fig:suba}
		\includegraphics[width=.45\textwidth]{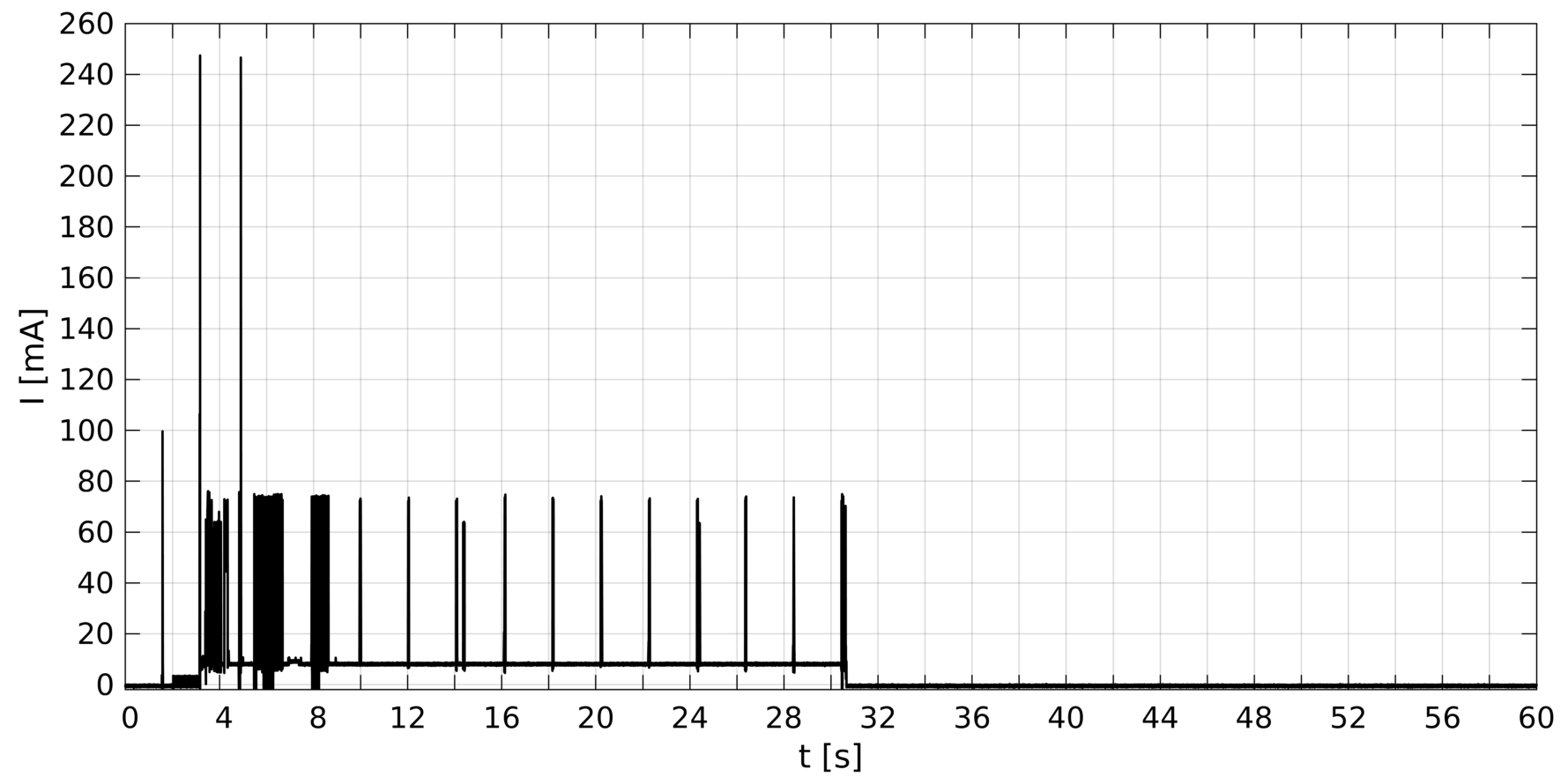}}
		%\quad
     \subfloat[][The connection is immediately released after sending the datagram (No C-DRX). 
     The module continues listening in I-DRX for 20 seconds. 
     This value is controlled by the Active Timer.]{%
        \centering
		\label{fig:subb}
	    \includegraphics[width=.45\textwidth]{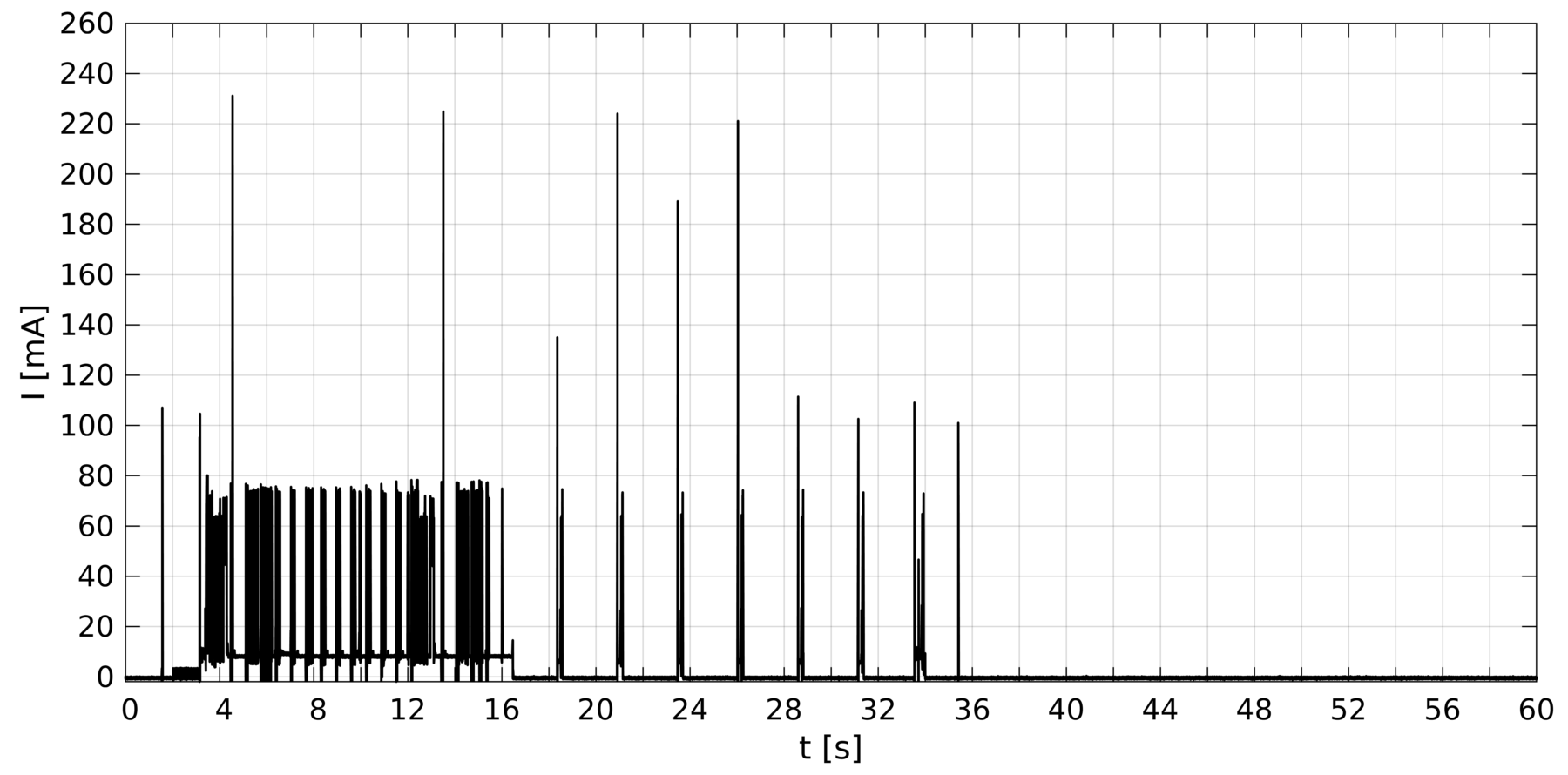}}
	    \quad
	\subfloat[][The connection is immediately released after sending the datagram, 
	thus preventing C-DRX cycles of idle listening. 
	I-DRX is also disabled by setting the Active Timer to zero.
	This setting provides minimal power.]{%
        \centering
        \label{fig:subc}
		\includegraphics[width=.45\textwidth]{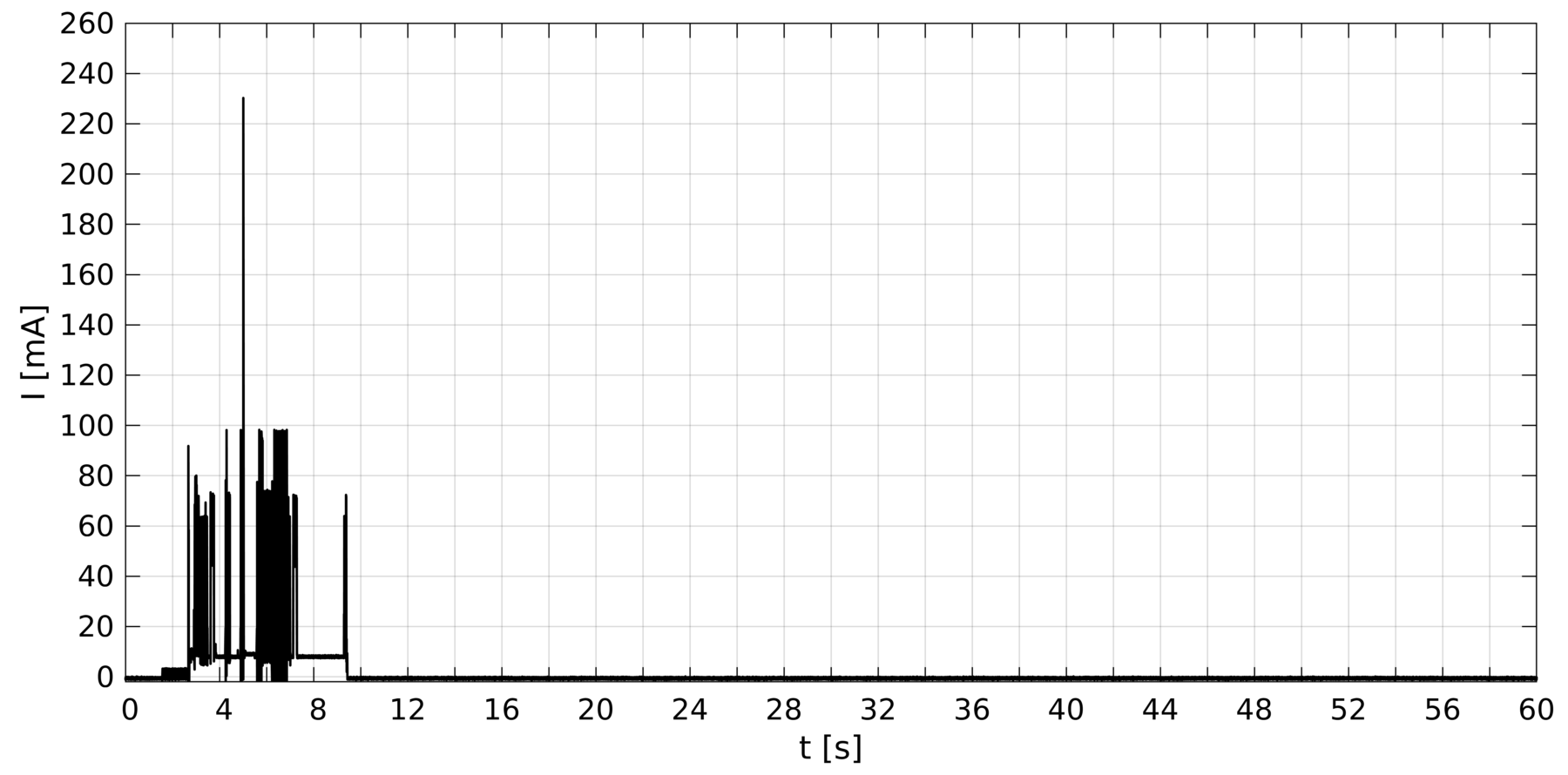}}
	\subfloat[][Same settings as in figure above.
	In this case, a downlink message is noticed while listening in I-DRX. 
	When the message is downloaded, it triggers a C-DRX cycle until Inactivity Timer expires.]{%
        \centering
		\label{fig:subd}
        \includegraphics[width=.45\textwidth]{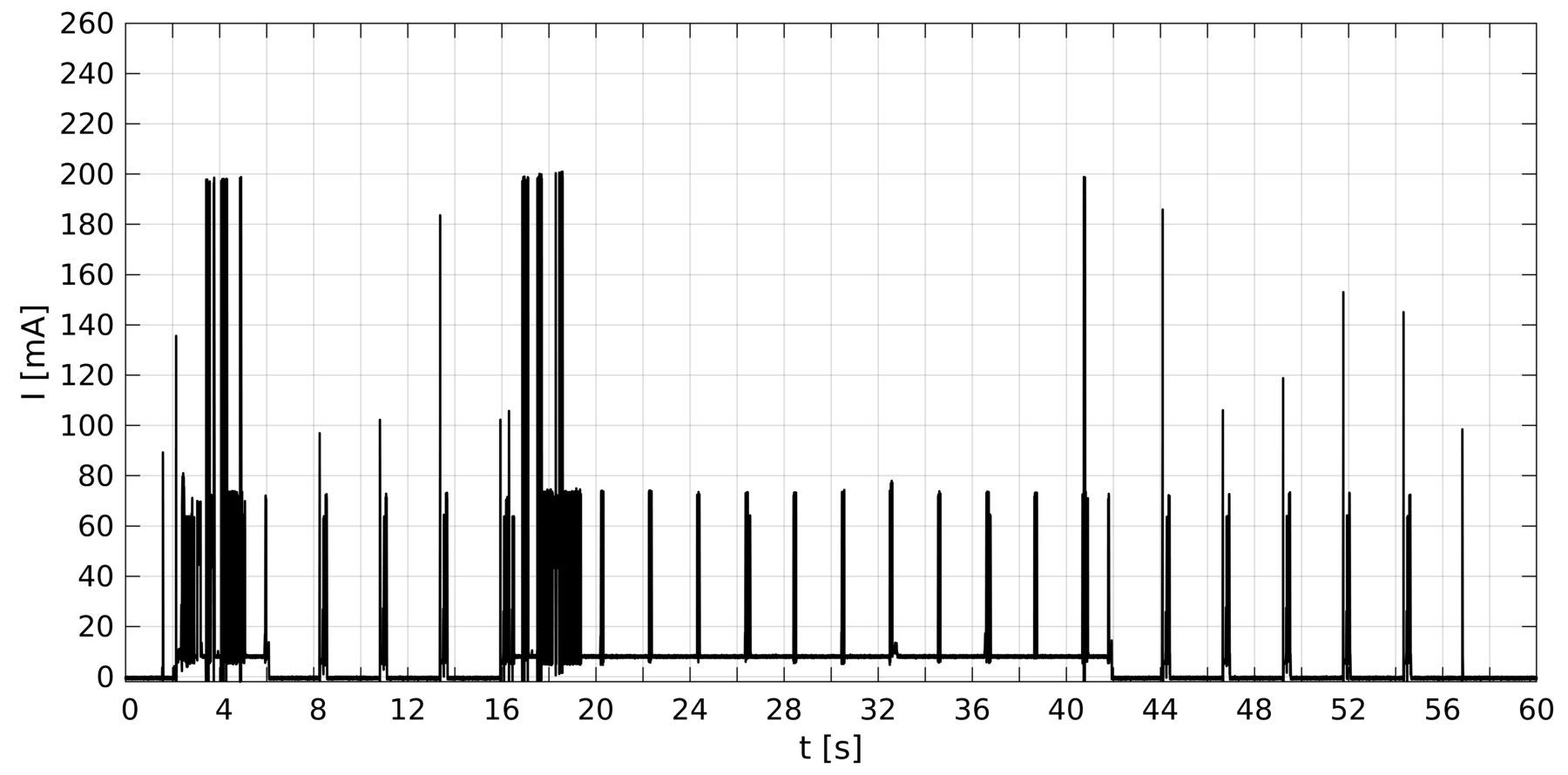}}
		\quad      
  \caption{Current traces of the UE sending a datagram of 512\,bytes with different network settings.}
  \label{fig:observ}
\end{figure*}

From a practical perspective, any NB-IoT chipset can be configured through a provided Application Programming Interface (API), typically in the form of a set of AT commands. This API is standardized by the 3GPP consortium as the \quotes{AT command set for User Equipment} \cite{3gpp.27.007}. 
Although different vendors may extend it with particular commands or shortcuts, all NB-IoT modules should be manageable through this standardized API. In general, an application developer has access only to the configuration options available through the API. This is important because even though the NB-IoT standard has been designed with multiple configurable options, and many articles discuss methods for ascertaining optimal settings, the application developer has actual control only for reaching a subset of operating points. The optimal settings methods may lead to suboptimal outcomes for the specific UE and its matching with the network, depending on what level of configurability the network supports. In this section we observe the energy signature of a UE using different configurations and we discuss network dependencies, especially those that are unable to be controlled from the application side.

\fig{fig:observ} presents measured current traces obtained from the \hide{Ublox SARA-N211 chipset}\footnote{To help identify the points of operation, the chipset features 3\,$\mu$A in deep sleep state, 10\,mA in idle, 60\,mA when the radio is active in reception mode, and a variable value ranging from 60\,mA to 220\,mA in TX mode.}
when operating in the NB-IoT Vodafone Network.
In \fig{fig:suba} we can observe the UE resuming the RRC connection and transmitting a 512-byte UDP datagram.
After the data transmission stage, when there is no pending traffic in any direction, the UE enters into Connected-mode DRX (C-DRX), where it monitors the control channel at regular intervals. The figure shows the peaks of the radio in reception mode during the wake-up cycles. These peaks are spaced at 2.048\,s intervals, which correspond to the 2048 subframes that are defined as one of the discrete values in the standard \cite{3gpp.36.331}.
%\todo{[ETSI TS-136.331, drx-Cycle-r13]}.
The network Inactivity Timer expires after 20\,s of inactivity, and the UE enters PSM mode afterward. Both the Inactivity Timer and the C-DRX are configured by the Vodafone network and cannot be changed from the UE. Unexpectedly, this particular chipset remains in idle mode (\texttildelow10\,mA) in the intervals between radio peaks, thus wasting any opportunities to enter deep sleep.

\fig{fig:subb} presents the trace of an equivalent transmission, but after configuring the UE for immediate release of the connection. In this case, the T3324 timer is configured so that the UE remains in Idle-mode DRX (I-DRX) for 20\,s before entering PSM. In idle state, DRX cycles occur every 2.56\,s. This is one of the four default paging intervals defined in the standard, which corresponds to 256 radio-frames \cite{3gpp.36.331}.
%\todo{[ETSI TS-136.331; defaultPagingCycle-r13]}.
Although the UE can request other tentative values, the network did not accept any other configuration in our tests. 
Finally, it is worth mentioning that,
unlike in the previous case with RRC Connected, in RRC Idle the chip enters deep sleep between radio cycles, reducing the current drain to a few microamps.

In \fig{fig:subc} we observe an uplink transmission with immediate release and the T3324 timer set to zero (I-DRX disabled). With these settings, the UE enters PSM directly after the datagram is sent and the RRC Connection is released. This example therefore presents the most basic use-case, that is, one in which the uplink is used to send a single datagram while the downlink is basically ignored. Despite this simplistic use-model, the complexity inherited from the underlying LTE network is noticeable in the figure, especially when compared with random access technologies such as LoRaWAN or SigFox 
(see \cite{Power2015,Raza2017LPWAN}, for example).

Finally in \fig{fig:subd} we observe an uplink and downlink sequence. There, the UE transmits and immediately releases the connection. Active Timer (T3324) is set so that the UE goes to RRC Idle state after the transmission, wherein it continues to monitor the control channel in I-DRX mode. Before the timer expires approximately 8\,s after the connection release, the server uses some paging occasion to send notification that downlink data is addressed to the UE. This event triggers the RRC Resume mechanism to receive the 16\,bytes datagram. Following the reception, the UE waits for the Inactivity Timer to expire (20\,s on this network), and the connection is released again to RRC Idle. 
Finally, the UE waits for the Active Timer to expire (16\,s in this example) before entering PSM.

From the above observations we conclude:
%\begin{enumerate}
\begin{enumerate}[label=\roman*)]
	\item The variability in terms of network activity and, therefore, 
	the required time for transmission of a single datagram is noteworthy. 
	This may have a significant impact on energy consumption, especially regarding predictability.
	\item The Inactivity Timer cannot be controlled from the UE. Depending on the network configuration, this can be a major energy-saving issue, especially if the chip is not able to enter deep sleep during the paging idle intervals, as in the analyzed example.
	\item The T3324 Timer is reset after a downlink message is received. 
	The negative impact on energy savings should be taken into account if downlink data is fragmented.
	\item {The transmission power varies dynamically because it is adjusted internally by the UE according to the allocated bandwidth and path loss (refer to \cite{3gpp.36.213} for further details on uplink power control). The control of this feature is not exposed through the API and therefore this may cause differences for apparently identical transmissions.} 
	This behavior can be observed for example in \fig{fig:subc}, on which we observe transmission peaks ranging from 100mA to 220mA approximately.
	\begin{comment}
	\item The transmission power varies dynamically because it is adjusted by the eNB according to the link quality of the received packets (refer to \cite{3gpp.36.213} for further details on uplink power control). 
	This may cause differences for identical transmissions. 
	An example of this behavior can be observed in \fig{fig:subc}, where we can observe transmission peaks ranging from approximately 100\,mA to 220\,mA.
	\end{comment}
\end{enumerate}

As a final remark, it should be taken into account that configurable options at the application level focus on optimizing the downlink operation of a UE. In contrast, the UE has very few options for optimizing the uplink operation, which implies that energy expenditure is mandated almost exclusively by the state of the network.

%-----------------------------------------------------------------------
% 4.- Performance
%-----------------------------------------------------------------------
% \input{nbiot-sec-performance}
% ======================================================================

\section{Performance Analysis}
\label{sec:performance}
In this section, we explore the performance in terms of the energy consumption of NB-IoT technology. Other functional aspects are also discussed, such as latency and reliability. We are particularly interested in finding the operational boundaries of NB-IoT, a goal that we address through a comprehensive data record of close to 3000 traces, which we believe may serve as a good representation of what an adopter can expect in the long term from this technology. Each trace, like those shown in \fig{fig:observ}, includes the resume process of the RRC connection, the actual transmission, the RRC release, and the subsequent transition to PSM.

In order to provide unbiased results, two commercial platforms from different vendors are used for the evaluation: \hide{: 
Quectel BC68\footnote{\url{https://www.quectel.com/product/bc68.htm}} 
and Ublox SARA-N211\footnote{\url{https://www.u-blox.com/en/product/sara-n2-series}}}. 
The experiments were conducted using the Vodafone NB-IoT Network (band 20), which is deployed in the metropolitan area of Barcelona.

The tests were designed to reproduce as closely as possible the IoT model described in the first section, for which smart-metering is our reference example. In this model, communications are always initiated from the UE to periodically report small chunks of data. The (occasional) downlink communications are scheduled as responses to these transactions, for which the UE opens a small listening window after each transmission. The device enters deep sleep (PSM) during idle periods between transactions.

Three different UE and network configurations are used throughout the study.

\begin{itemize}
\item Mode 1: The listening window opens in RRC Connected mode. 
The duration is determined by the Inactivity Timer, 
which is managed by the network.
\item Mode 2: The RRC connection is released immediately after transmission, and the listening window opens in RRC Idle. The duration is determined by the Active Timer of the device and is therefore configurable.
\item Mode 3: The connection is released immediately after transmission.
DRX is disabled (no listening window), so communication is basically unidirectional.	
\end{itemize}
The specific settings for these configurations are summarized in \tab{tab:config}.
     
\begin{table}[h!]
\centering
\caption{Summary of the evaluated configurations}
\label{tab:config}
\begin{tabular}{|l|l|}
\hline
	\textbf{Mode 1} & 
	\begin{tabular}[c]{@{}l@{}}
	Inactivity timer = 20\,s (network default)\\ 
	T3324 = 0\,s (disabled) \\
	C-DRX = 2.048\,s (network default)
	\end{tabular} \\
\hline
	\textbf{Mode 2} & 
	\begin{tabular}[c]{@{}l@{}}
	Inactivity timer = Immediate Release\\ 
	T3324 = 8\,s \\ 
	I-DRX = 2.56\,s \\ 
	eDRX/PTW = Disabled 
	\end{tabular} \\
\hline
	\textbf{Mode 3} & 
	\begin{tabular}[c]{@{}l@{}}
	Inactivity timer = Immediate Release\\ 
	T3324 = 0\,s (disabled)
	\end{tabular}\\
\hline
\end{tabular}
\end{table}

%Along this analysis we will observe 
In our experiments, different values of SNR were forced by means of attenuators physically connected to the antenna. The objective was to enforce the mechanisms designed in NB-IoT to improve coverage, mainly repetitions and variable transmission power. It is worth noting that the coverage must be guaranteed by the network operator through proper network deployment planning. Therefore, the performance of NB-IoT must be evaluated from an adopter/user perspective under different coverage conditions, that is, under different ECL.

We also stress that many features of NB-IoT are beyond the control of the user, especially in the uplink: the transmission power and repetitions are negotiated between the UE and the eNB, while retransmissions depend entirely on the state of the network. Therefore, we rely on a probabilistic analysis after collecting a large number of samples.

{Finally, it should be noted that transmission mode optimizations such as UP and CP are beyond the control of the 
end user for the platforms under evaluation. Therefore the results derived from our experimentation campaign are conditioned to the mode selected internally by the UE.}

%-----------------------------------------------------------------------
% Uplink Energy
%-----------------------------------------------------------------------

\subsection{Overall Behavior}
\label{subsec:iv-1}

\begin{figure}[!h]
  \centering
  \includegraphics[width=1.0\columnwidth]{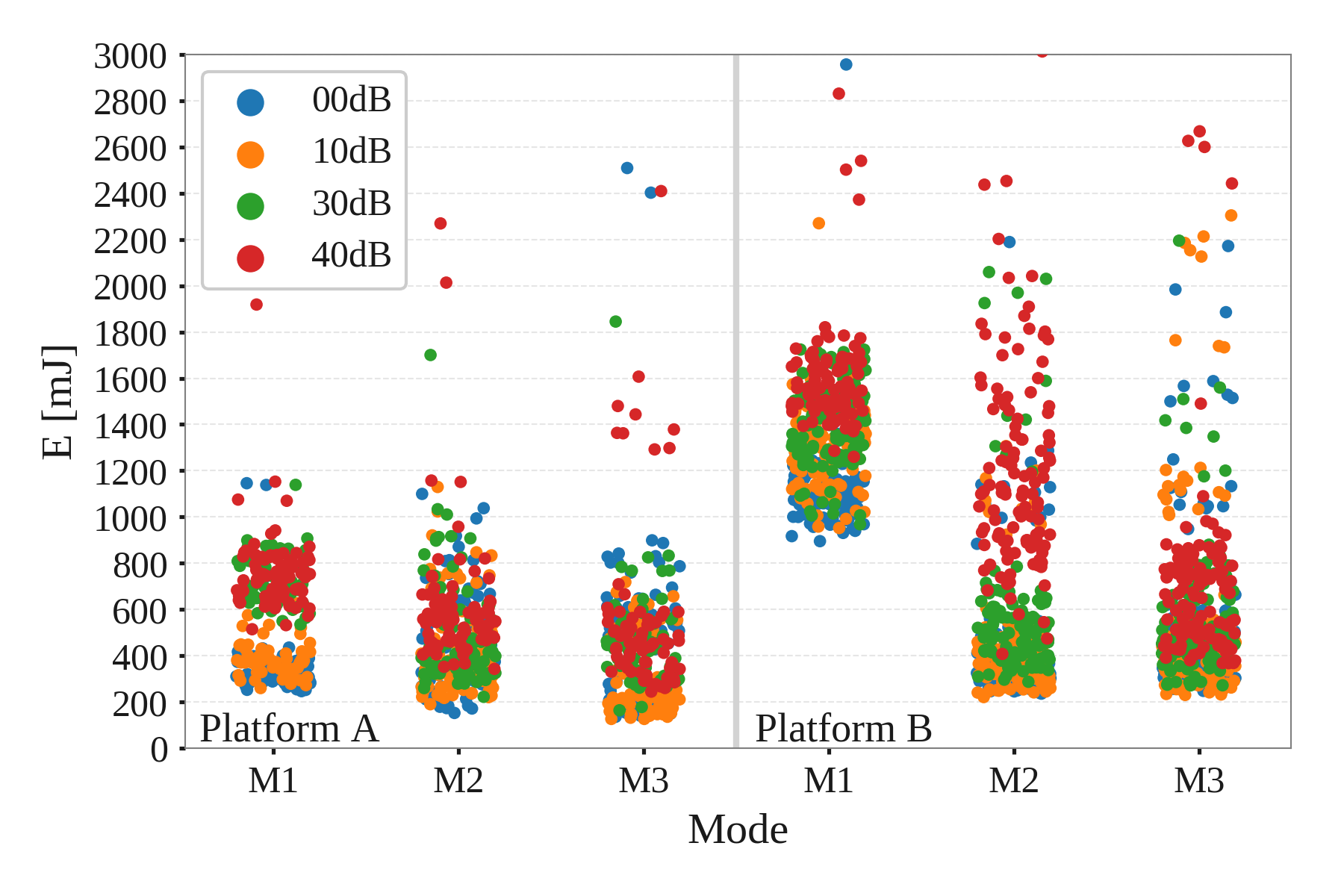}
  \caption{Energy consumed when sending a single datagram using the three network settings defined in \tab{tab:config} for each of the tested platforms\hide{, Quectel and Ublox}.}
  \label{fig:energy-mode}
\end{figure}

A first analysis aims to understand the overall energy consumption of the UE, according to the selected mode of operation and when subject to different signal quality scenarios. The analysis also evaluates the two platforms in order to understand if some differences can be attributed to the particular hardware. 
\fig{fig:energy-mode} presents the corresponding energy consumption per UDP datagram sent, showing the performance for the two devices \hide{(Quectel and Ublox)} and for the three evaluated modes (\tab{tab:config}). 
The energy is obtained by integrating the current trace measured during the transaction at the voltage supply (3.3\,V). All results are labeled according to the attenuator used for that particular record.

From the figure, we can make some observations:
%
%\begin{enumerate}[label=(\alph*)]
\begin{enumerate}[label=\textbullet]
\item Mode 1 incurs more energy than the others. This is explained mainly by the UE listening for the NPDCCH during the Inactivity Timer period (20\,s) every 2.04\,s. In this mode, there is a significant difference between the 
two vendor platforms, as the \hide{Ublox} platform does not seem to enter deep sleep during idle states of RRC Connected. This limitation can be attributed exclusively to the firmware/hardware, and therefore is not relevant in our study.

\item Modes 2 and 3 perform similarly, even though Mode 2 enables downlink for 8\,s after an uplink window while in RRC Idle. 
From this fact we can infer that idle listening has little impact, at least when the listening window is small.
\item We observe that there is a slight energy increase as the received signal strength decreases. We attribute the variation to the higher transmit power and higher number of repetitions when signal quality is lower. 
\end{enumerate}

% Power vs Payload
 
\begin{figure}[ht!]
  \centering
  \includegraphics[width=1.0\columnwidth]{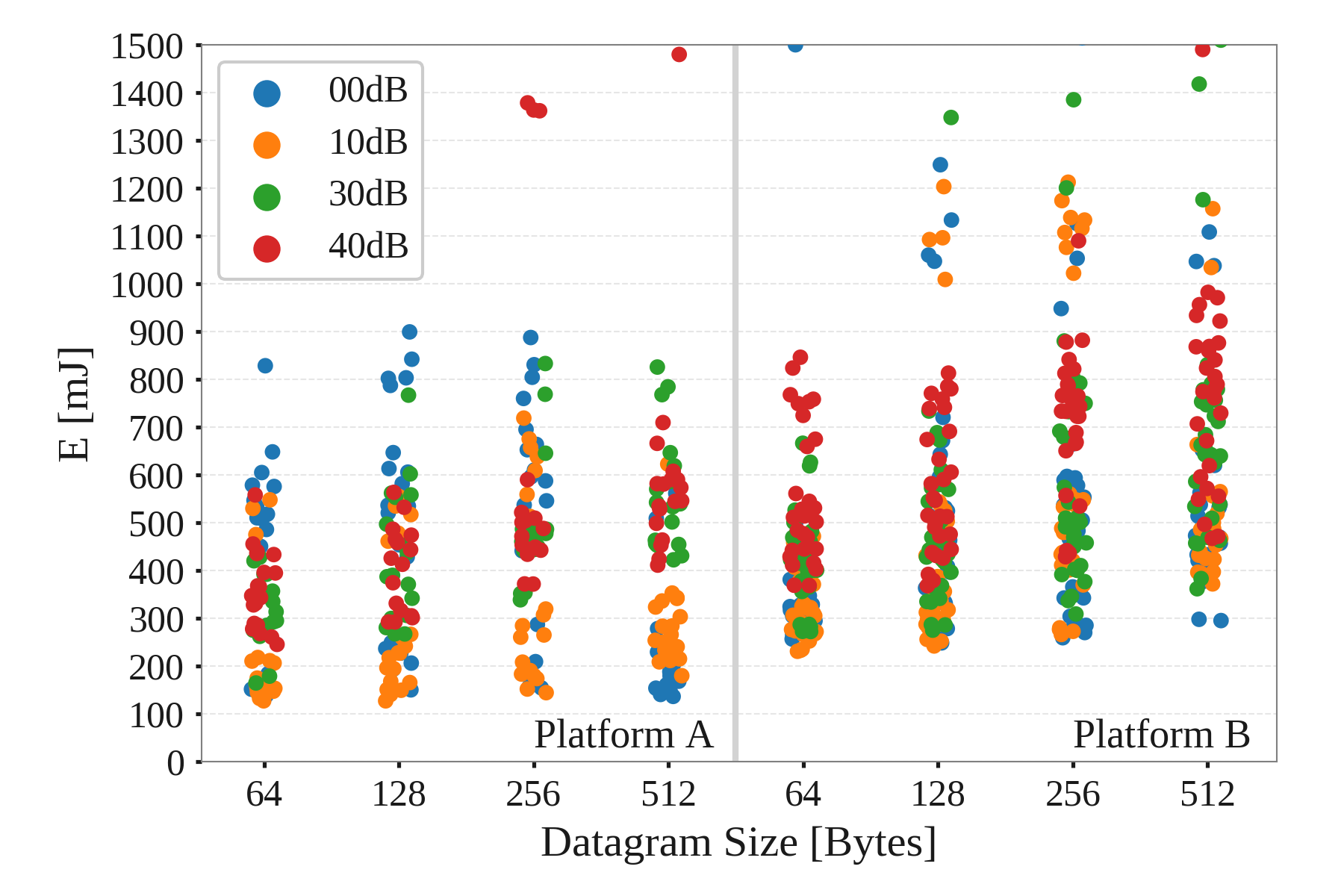}
  \caption{Energy expenditure per datagram by payload size for each platform\hide{, Quectel and Ublox}.}
  \label{fig:energy-size}
\end{figure}

\subsection{Uplink Power}

Fig.~\ref{fig:energy-size} shows the impact of payload size on energy consumption. 
In the figure, we can observe two main findings: 
%\begin{enumerate}[label=(\alph*)]
\begin{enumerate}[label=\textbullet]
\item Multiplying the packet size by eight barely increases energy consumption.
\item Given a particular packet size and even after fixing the UE's attenuation conditions, there is huge variability in energy consumption.
%\item As larger the packet the more relevant is the impact of smaller SNRs into the energy consumption.
\end{enumerate}

%-----------------------------------------------------------------------
%-----------------------------------------------------------------------
\begin{figure}[ht!]
  \centering
  \includegraphics[width=1.0\columnwidth]{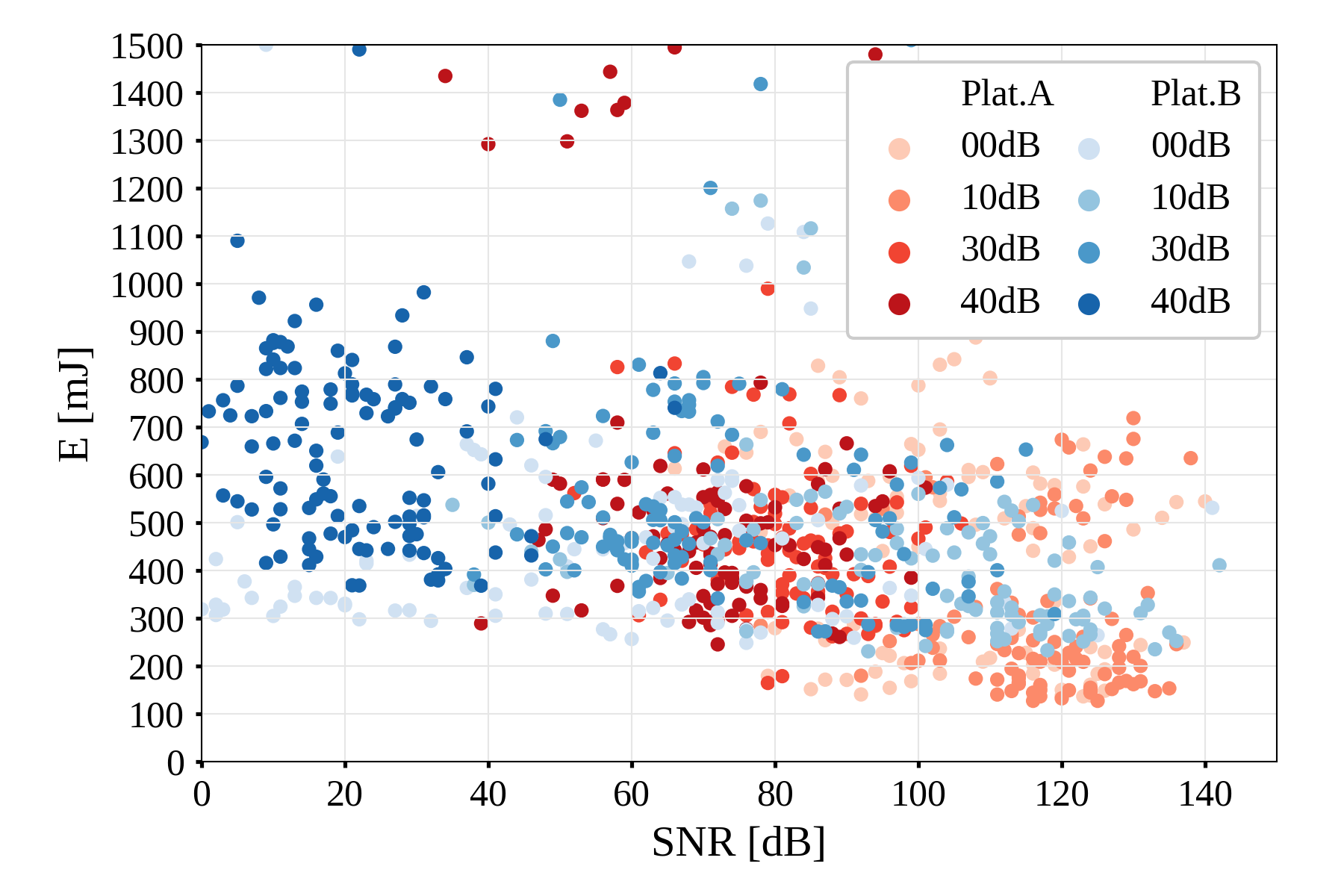}
  \caption{Energy per datagram as a function of the SNR reported by the UE after completing transmission.}
  \label{fig:energy-snr}
\end{figure}

In turn, \fig{fig:energy-snr} shows the impact of the SNR on the UE's energy consumption.  
As observed in the figure, the lower the SNR, the higher the energy consumption. This is mainly because coverage extensions enter into play. 
From our perspective, the most relevant observation is that there is a notable variability in energy consumption due to the SNR perceived by the particular UE. 
From an integrator perspective, this must be taken into account when dimensioning the device battery,
because according to the UE's environment (i.e., the coverage level) can reduce battery life expectancy to one-half or less.

%-----------------------------------------------------------------------
% Downlink 
%-------------------------------------------------------------------------------

\subsection{Downlink power and impact}
\begin{figure}[ht!]
  \centering
  \includegraphics[width=1.0\columnwidth]{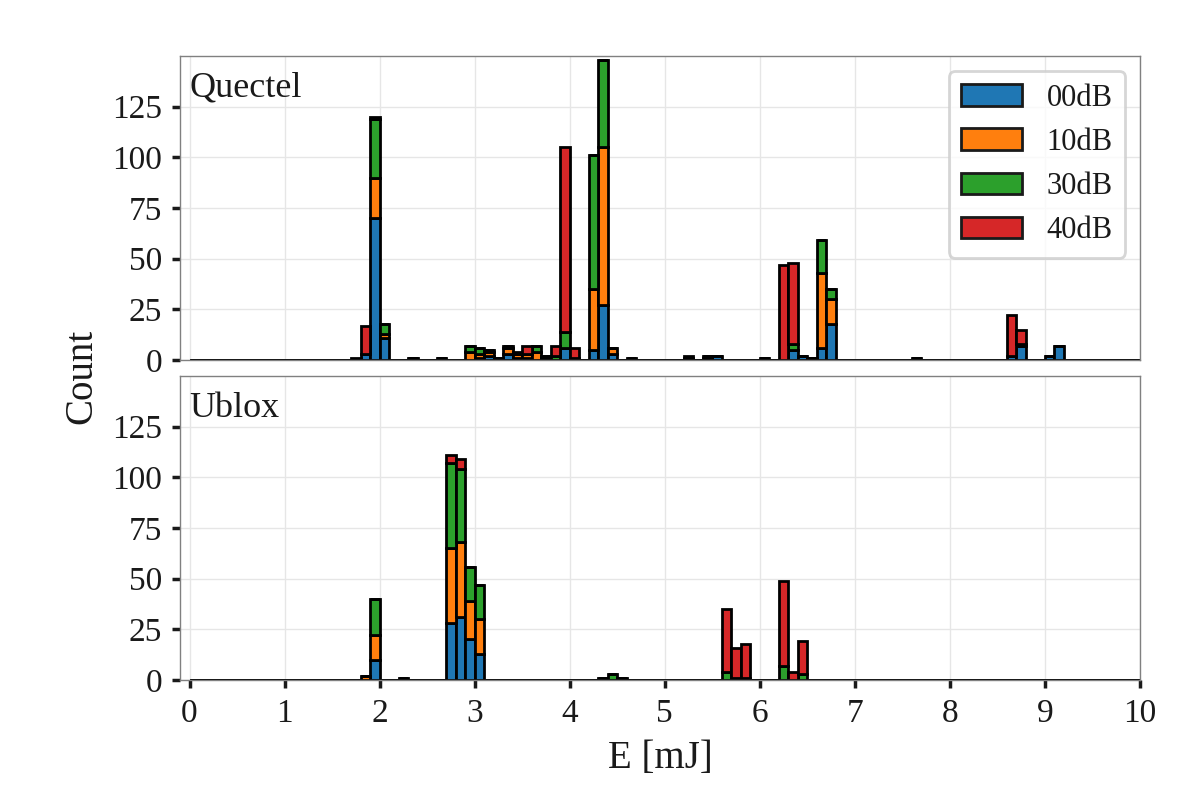}
  \caption{Histograms of the energy corresponding to the peaks produced by listening for paging occasions, compiled in the set of all tests.}
  \label{fig:energy-edrx-peaks}
\end{figure}

Downlink is enabled through eDRX cycles. UEs listen for paging occasions in order to detect if any downlink packet is queued to be downloaded.  
As we advanced in Section \ref{subsec:iv-1}, the cost of listening for paging occasions was small because NB-IoT duty cycles the subframes to which the radio is listening. 
In \fig{fig:energy-edrx-peaks} we can observe a histogram of the energy required to listen one paging frame in I-DRX. 
The bars indicate the number of occurrences of such energy from the whole set of experiments in our data set. 

From the figure, we can derive the following conclusions:
%\begin{enumerate}
\begin{enumerate}[label=\textbullet]
	\item The energy required to track POs is two orders of magnitude smaller (in the order of mJ) when compared to the energy required to send a message (on the order of hundreds of mJ, see \fig{fig:energy-size} for a reference). 
	\item We cannot conclude there is a direct relationship between the energy required to track POs and the signal attenuation at the UE.
	While \hide{Ublox} transmissions with maximum attenuation seem to be clustered with higher energy, \hide{Quectel} data does not exhibit the same behavior.
	\item The energy peaks are grouped around discrete values, and this seems to be related to the number of repetitions. 
\end{enumerate}

In general terms, compared to the cost of transmitting a packet, a nearly insignificant cost is incurred from enabling downlink capabilities for a short period after an uplink window.

%-----------------------------------------------------------------------
% Delay
%-------------------------------------------------------------------------------

\subsection{Delay}
\begin{figure}[ht!]
  \centering
  \includegraphics[width=1.0\columnwidth]{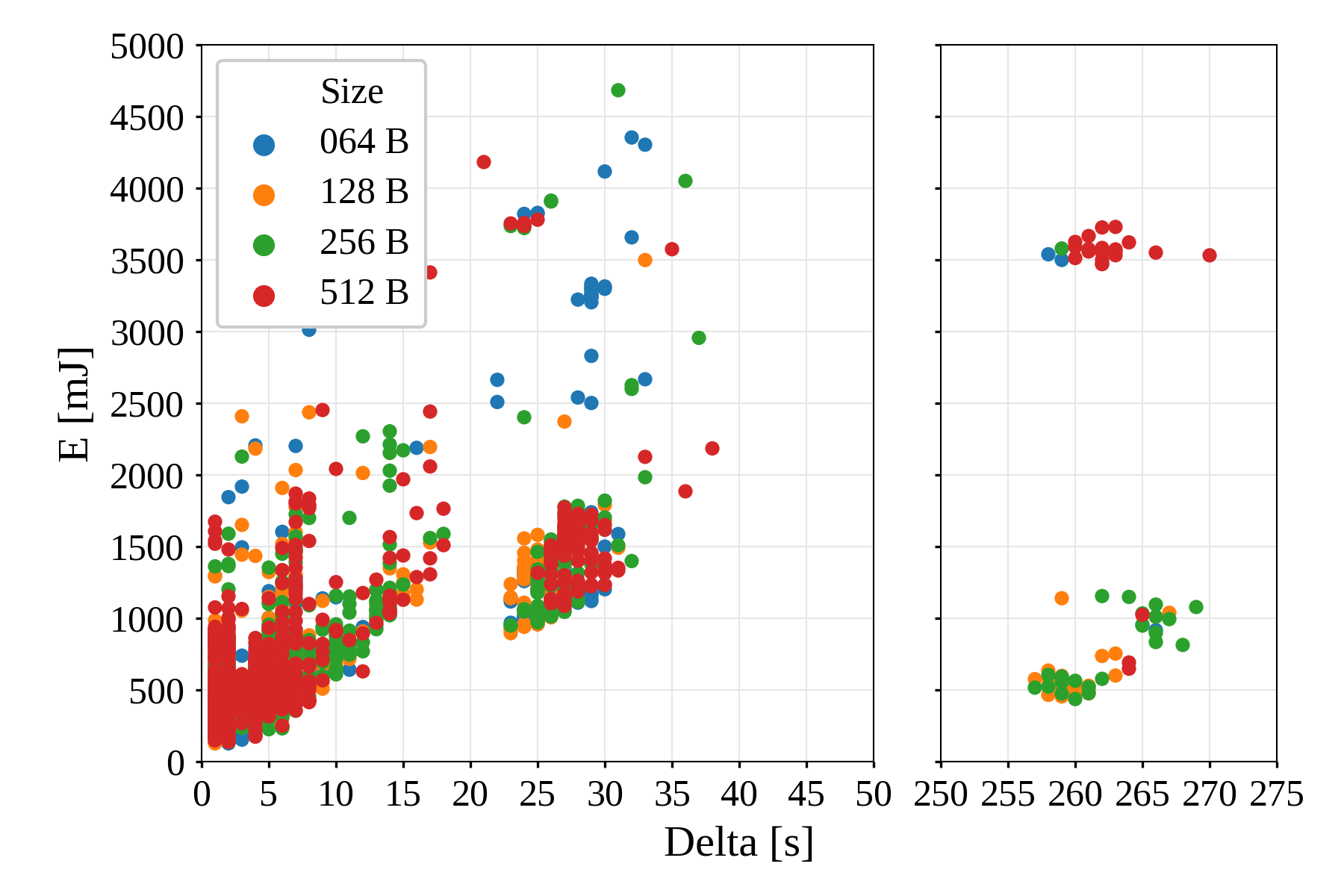}
  \caption{Delay of arrival to the server and its relationship to the energy measured on the device. 
  The dots are split by colors according to the enforced attenuation.}
  \label{fig:energy-delay}
\end{figure}

NB-IoT has been designed for delay-tolerant applications. 
In our study, we aimed to empirically analyze the network delay under the configurations described above. 
\fig{fig:energy-delay} presents the measured delay of 2880 UDP datagrams transmitted during our tests. 
The delay is obtained as the difference between the transmission time at the UE and the reception time at an Internet reachable server located at our premises. 
It should be noted that datagrams are sent independently, 
so each transmission requires that the RRC resume process to be executed before.
The resume time is included in the delay reported.

From \fig{fig:energy-delay} we can observe:
\begin{itemize}
	\item The delay is not dependent on the message size.
	\item The greater the delay, the greater the energy consumption. This is explained by the fact that the UE is waiting for the message acknowledgment. 
	\item Three delay regions appear in our analysis. The first region includes datagrams that took between 0\,s and 18\,s to reach the destination (2307/2880). The second region corresponds to datagrams that took between 21\,s and 39\,s (496/2880). A third region groups datagrams that took between 256\,s and 270\,s (77/2880). 
\end{itemize}

\begin{figure}[ht!]
  \centering
  \includegraphics[width=1.0\columnwidth]{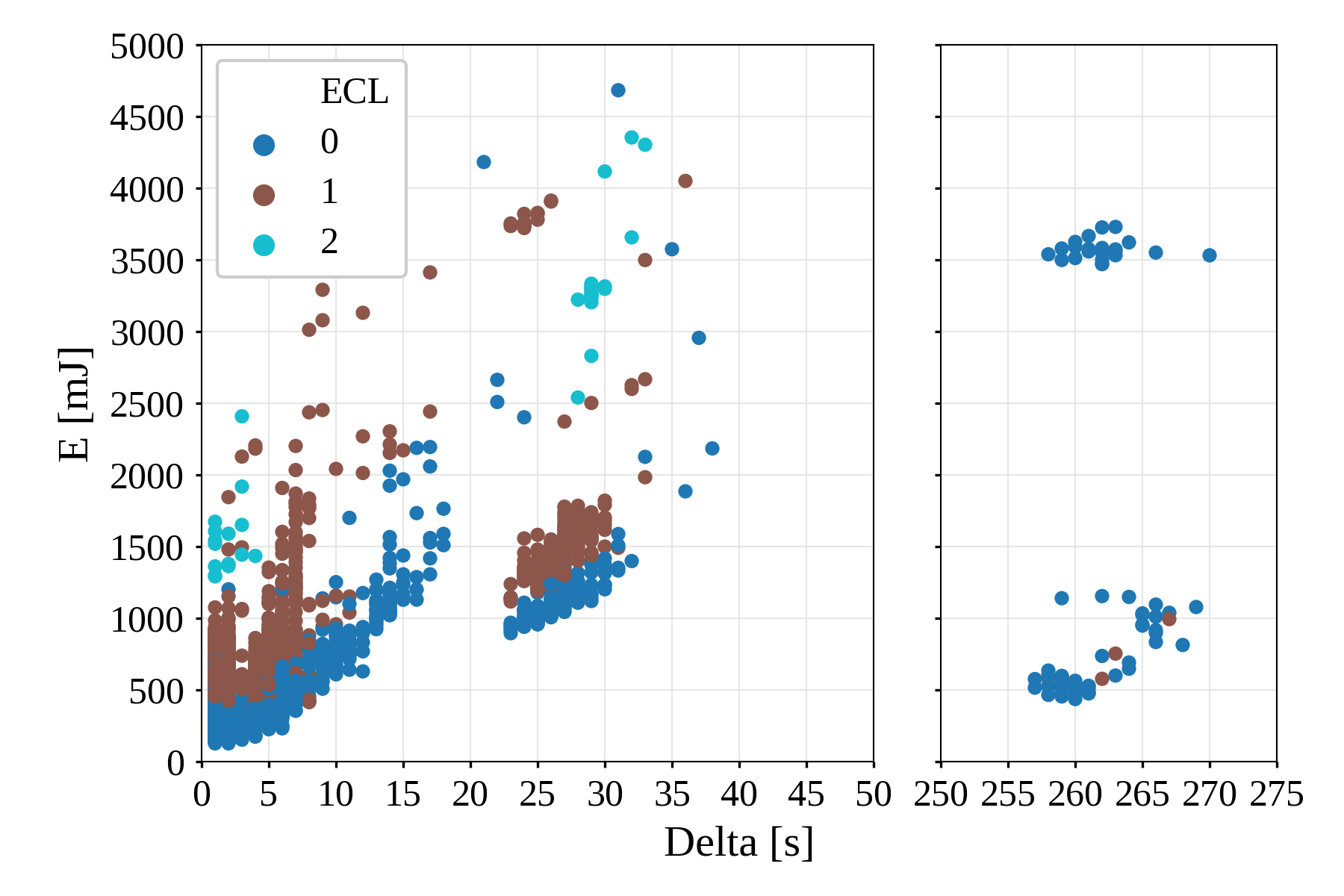}
  \caption{Delivery delay and its relationship to the energy measured. 
  The dots are split according to the reported ECL after transmission.}
  \label{fig:energy-delay-ecl}
\end{figure}

One might think that these regions can be mapped to the different ECLs supported by the network. \fig{fig:energy-delay-ecl} disproves this speculation. As observed, the ECL has an impact on energy consumption, but not on the delay. 

\subsection{Final remarks}

Based on the above analysis and taking an adopter perspective, we would like to emphasize some observations. 
First, the combination of two factors, namely, 
the accommodation within LTE (signaling overhead) and coverage enhancements (e.g. repetitions), 
generates a complex behavior with high variability. 
This variability is reflected in the energy consumption,
and results in poor predictability of battery life, thus causing a divergent behavior between similar devices.
This is the price to pay for guaranteed reliability in NB-IoT,
and it must be taken into account for applications where lifespan forecast is critical, such
as those subject to service-level agreement (SLA).  
Second, although NB-IoT is designed for delay-tolerant applications,
there are cases in which dalays of tens of seconds or even minutes, may not be acceptable.
Finally, as in LTE networks, application developers must be aware that the described variability
is not under their control.

%-----------------------------------------------------------------------
% 5.- Comparative analysis: NB-IoT vs LoRa
%-----------------------------------------------------------------------
%\input{nbiot-sec-positioning}
% ======================================================================

\section{NB-IoT positioning}
\label{sec:comparative}

\begin{figure}[t!]
  \centering
  \includegraphics[width=1.0\columnwidth]{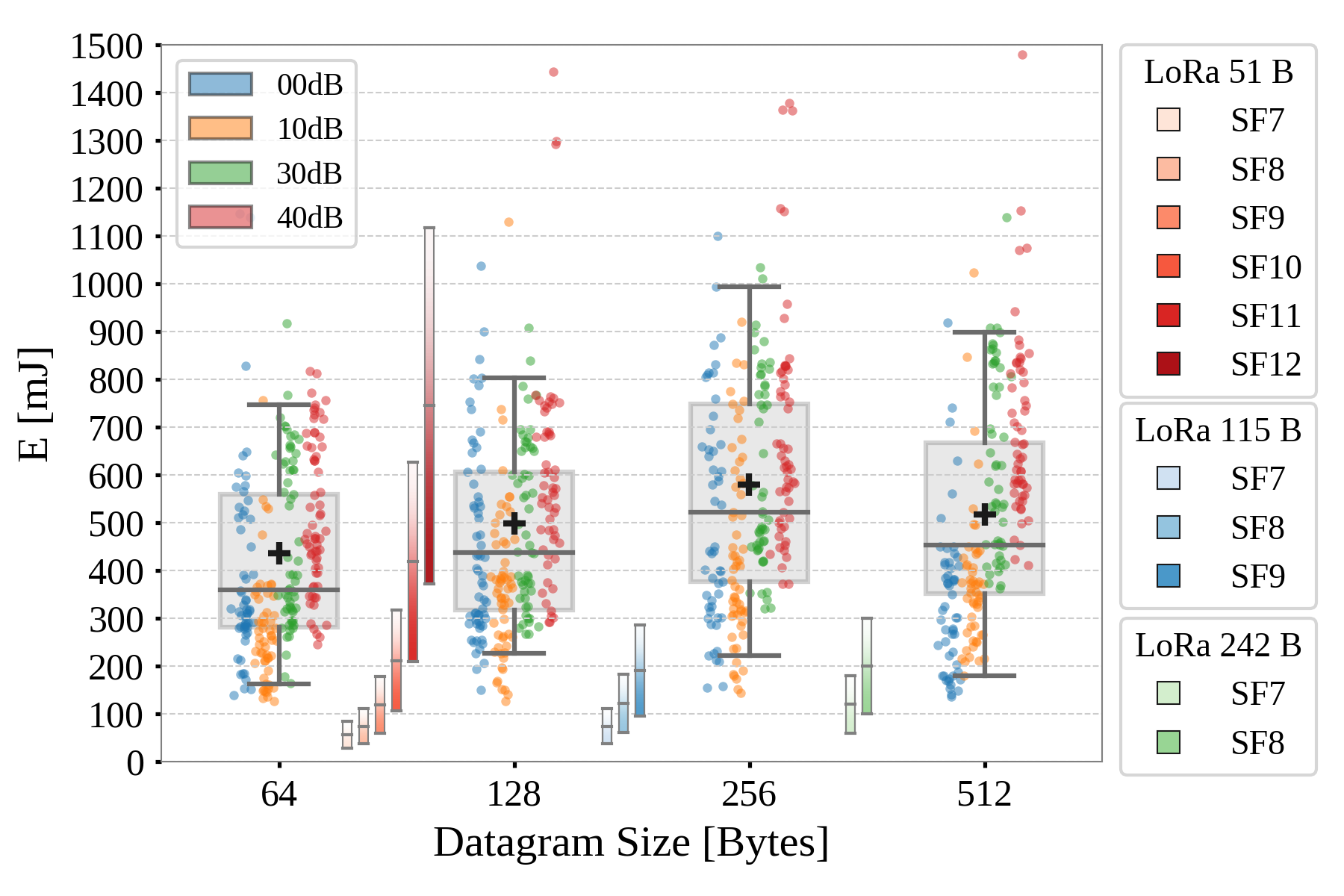}
  \caption{Boxplot representation of NBIoT energy traces compared to LoRaWAN. 
  The median, first and third quartiles are depicted with the box. 
  The whiskers indicate 5th-95th percentiles and the black cross the mean value.
  LoRaWAN values are depicted for different spreading factors and size.}
  \label{fig:lora-nbiot-energy}
\end{figure}

NB-IoT is a 3GPP response to the demand for LPWAN technologies, and it is beginning to compete with well-established technologies on the market, such as LoRaWAN, Wireless M-BUS and Sigfox (see \fig{fig:e1}). Amongst them, LoRaWAN is the most adopted technology because it enables ad-hoc, simple deployments that support small- to large-scale networks without involving an operator. Moreover, LoRaWAN-operated networks are possible through different service providers, including open and free communities such as The Things Network. Details about the performance and limits of LoRaWAN can be found in \cite{Adelantado17Lorawan}. 
Being that LoRaWAN is a major contender, our goal is to compare NB-IoT to it.

\subsection{Energy per message}

\fig{fig:lora-nbiot-energy} compares the energy that both technologies require to transmit an application layer message. In the case of NB-IoT results, we use the experimental results described in the previous section. As already mentioned, NB-IoT is subject to high variability in terms of energy consumption. To account for this variability, a box plot is used to depict the median, first and third quartiles. The whiskers indicate the 5th-95th percentiles. The mean value is included for completeness and is marked with a black cross, as the distributions are not symmetrical. 

However, despite of this high variability, NB-IoT guarantees message delivery. 
Layer 2 reliability mechanisms enable an application to rely on the network infrastructure to ensure delivery. 
In opposition, Aloha-based LPWANs in general, and LoRaWAN in particular, are constrained in the downlink, due to duty-cycle regulations in the Industrial Scientific and Medical (ISM) band. 
Since sending an acknowledgment message through the downlink channel for all messages is not possible, 
users are forced to develop their own strategies (e.g., repetitions), whose impact is difficult to quantify. 
For example, SigFox makes 3 retransmissions \cite{Power2015} thus increasing the power accordingly.
Even so, delivery is not guaranteed.
In addition, Time Division Multiple Access approaches for LoRaWAN have been explored in the literature \cite{Reynders2018}.
These approaches need to deal with the duty-cycle limitations 
and, therefore, their performance and scalability are quite limited.
%LoRaWAN energy is obtained following the model defined by Casals et.al~\cite{Casals2017}.

These fundamental differences between both technologies make it difficult to compare them fairly.
%The energy required to transmit a LoRaWAN message is calculated for each of the spreading factors (SF7-SF12) 
In \fig{fig:lora-nbiot-energy}, we depict the LoRaWAN consumption with vertical bars. 
The energy required to transmit a LoRaWAN message has been measured for each of the spreading factors (SF7-SF12) 
while considering the maximum supported packet sizes depending on the SF (51\,B, 112\,B and 251\,B)%
\footnote{In our experiments we used the B-L072Z-LRWAN1 platform, 
values for an alternative platform (LoRaWAN Multitech mDot) can be found in \cite{Casals2017}.
Complementary work can be found in \cite{Bouguera2018}.}.
Then, for each bar, lower limit corresponds to the transmission of each packet only once, 
with neither retransmissions nor repetitions, so the delivery is not guaranteed. 
This would therefore be the lower bound of LoRaWAN in the most idealistic scenario. 
In turn, the upper limit of each bar corresponds to 3 retransmissions of each packet (equivalent to Sigfox's strategy).
The exact value with which it is to be compared will therefore depend on the specific application (strategy) and the state of the network.

In the figure we can observe that for small packets, the mean energy required to send a NB-IoT packet is comparable to sending one LoRaWAN datagram at {SF11-SF12}. 
Larger packets can only be sent in the smaller SF which limits the range of LoRaWAN when compared to NB-IoT. Hence NB-IoT provides better coverage and network capacity for large packets.
For small payloads LoRaWAN allows the use of the highest spreading factors (SF11 and SF12), 
which implies a wider coverage. 
If in these cases some additional mechanism is required to guarantee delivery, we can see in the figure that LoRaWAN energy is not only comparable to the worst cases for NB-IoT, but also much higher than the average. 
 
\subsection{Application example}

In a simple periodic-reporting application with very limited computing requirements\footnote{Smart metering is a good example for which this simple model is valid},
the average power can be modeled approximately by \equ{eq:PnetPtoPperiodic}, as detailed in \cite{Power2015}:
 
\begin{equation}
  \bar{\MP} = \frac {\ME_{MSG}} {T_{MSG}}
  \label{eq:PnetPtoPperiodic}
\end{equation}

Periodic-reporting means that the time between messages ($T_{MSG}$ in \equ{eq:PnetPtoPperiodic}) can be considered a constant parameter.
However, due to the demonstrated energy variability of NB-IoT,
an estimate of the energy per message $\ME_{MSG}$ must be chosen in accordance with the application requirements,
ranging from very optimistic (best case) to the most pessimistic (worst case). 

For that purpose, we use the data recorded as a probabilistic model,
taking the 5th/95th-percentiles for the best/worst case scenarios,
and the mean values as an estimate for the long-term behavior.
The values obtained are compared to the best case setting using LoRaWAN.
\tab{tab:power} presents the average power of NB-IoT compared to the LoRaWAN lower limit 
(no retransmissions, no repetitions) when configured for reporting intervals of 1\,h. 
As can be observed, mean values for NB-IoT are similar to the energy that a LoRaWAN device requires to transmit while using the SF12 configuration. 
The 5th percentile results for NB-IoT (best observed performance) are comparable to the best case performance of LoRaWAN when operating at SF8. 
This is in our opinion a relevant result, as NB-IoT guarantees packet delivery with similar power consumption.

\begin{table}[]
\centering
\caption{Average Power Consumption}
\label{tab:power}
\setlength{\tabcolsep}{4pt}
\begin{tabular}{@{}c|ccc||ccccc|c@{}}

  \multicolumn{4}{c}{\textbf{NBIoT}}  & 
  \multicolumn{6}{c}{\textbf{LoRa}} \\ 
  \toprule
    Size & 5\% & Mean & 95\% & SF8 & SF9 & SF10 & SF11 & SF12 & Size  \\ 
  \midrule
	\ph{0}64 & 44 & 121 & 209 &     10  &      16  &     29   &      58  &     103  &  \ph{0}51 \\
         128 & 62 & 138 & 226 &     17  &      26  & \ph{58}  & \ph{116} & \ph{206} &       115 \\
         256 & 61 & 161 & 276 &     28  &  \ph{52} & \ph{116} & \ph{232} & \ph{412} &       242 \\
         512 & 49 & 143 & 250 & \ph{56} & \ph{104} & \ph{232} & \ph{464} & \ph{824} &           \\
   \midrule 
   \multicolumn{10}{l}{Power in [{\textmu}W]. Reporting interval T$_{\text{MSG}}$=1\,h.}\\ 
   \bottomrule
  \end{tabular}
\end{table}

%[0.01025    0.01650528 0.02938333 0.05805667 0.10349139]
%[0.01692528 0.02654361]
%[0.02776889]

\begin{table}[]
\centering
\caption{Estimated battery life}
\label{tab:lifespan}
\setlength{\tabcolsep}{4pt}
\begin{tabular}{c|ccc||rrrrr}

  \multicolumn{4}{c}{\textbf{NBIoT}} & 
  \multicolumn{5}{c}{\textbf{LoRa}}  \\ 
  \toprule
   N & 5\% & Mean & 95\% & SF8 & SF9 & SF10 & SF11 & SF12   \\ 
  \midrule
    \ph{0}64 & 8.4 & 3.1 & 1.8 & 29.3 & 18.2 & 10.2 & 5.2 & 2.9  \\
         128 & 6.0 & 2.7 & 1.7 & 19.5 & 12.4 &  5.1 & 2.6 & 1.5  \\
         256 & 6.1 & 2.3 & 1.4 & 12.8 &  6.2 &  2.6 & 1.3 & 0.7  \\
         512 & 4.5 & 2.6 & 1.5 &  6.4 &  3.1 &  1.3 & 0.6 & 0.4  \\
   \midrule 
   \multicolumn{9}{l}{Expected life in years per 1\,Ah at N\,[Bytes/h] (average)}\\ 
   
   %\multicolumn{9}{l}{Expected life in years per 1\,Ah of capacity. }\\ 
   %\multicolumn{9}{l}{Average throughput N\,[Bytes/h] }\\ 
                     
   \bottomrule
  \end{tabular}

\end{table}

% [29.3 18.2 10.2  5.2  2.9]
% [19.5 12.4  5.1  2.6  1.5]
% [12.8  6.2  2.6  1.3  0.7]
% [6.4 3.1 1.3 0.6 0.4]

\tab{tab:lifespan} calculates the expected lifetime for both technologies while considering %the same reporting interval (1\,h), assuming a $1Ah$ battery. 
an application that reports chunks of $N$ bytes per hour, assuming a $1Ah$ battery.
The expected achievable lifespan (on average) for a NB-IoT is on the order 2-3 years, depending on the datagram size. 
These values ​​are comparable to LoRaWAN, with SF12 sending an average of up to 64 bytes/h in messages of 51 bytes. 
However, adopters may take into consideration some differences.
First, sending larger messages (up to 512 bytes) has almost no impact on NB-IoT.
LoRaWAN is much more sensitive to the number of bytes to be sent. 
In general, LoRaWAN performs better for short messages, 
but it is subjected to a very high penalty when more than one message per data block is required.
Second, the LoRaWAN reliability mechanism must be ensured at the upper layers, 
and thus may incur higher energy costs. 
On the other hand, 
although the average power is comparable, 
peaks in transmission of LoRaWAN's radio are around 40\,mA, while in NB-IoT they reach 220\,mA. 
This causes additional stress on the battery, which has to be managed with care.

%-----------------------------------------------------------------------
% Conclusion
%-----------------------------------------------------------------------
% \input{nbiot-sec-conclusion}
% ======================================================================

\section{Conclusion}
\label{sec:conclusion}
In this article, we have evaluated the performance bounds of NB-IoT from an empirical perspective, which thus has allowed us to consider the application developer's position when adopting this technology. Such an approach aims to facilitate to the adopters the characterization of the application's behavior since them cannot control or parametrize all that is involved, such as signaling, dynamic adjustments triggered by the network conditions, and the timing that controls NB-IoT access.
NB-IoT has proven to be competitive in terms of energy consumption, which demonstrates the efforts made by the 3GPP to achieve the performance of other LPWAN technologies, even when realizing that the latter were designed from scratch with the main objective of optimizing power. Therefore, other features must be taken into account in order to choose the most suitable technology for each application. Among others:

%\begin{itemize}
%\item[$\to$]

\noindent$\to$\textit{Proprietary Spectrum:}
NB-IoT friendly coexists with LTE in a proprietary part of the spectrum. Technologies using ISM bands share the spectrum and may be subject to external interference. However, as we have seen when adapting to a cellular network structure, the complexity of the device’s behavior increases, which in the end leads to a high unpredictability

\noindent$\to$\textit{Reliability:}
The NB-IoT network guarantees delivery. This is an important aspect because alternatives like LoRaWAN can incur significant energy costs for guaranteed delivery, as they are also severely limited by duty-cycle regulations. If reliability is important, this can be a decisive factor.

\noindent$\to$\textit{Delay Tolerance:} The price to pay for low consumption in NB-IoT is high variability in delivery time. In our opinion, this can be one of the main deal-breakers in using NB-IoT for some applications.

\noindent$\to$\textit{Data rate:}
Most competitors in the LPWAN arena have been designed to transmit a few bytes per hour, even per day. If the application sporadically requires high bandwidth, NB-IoT may be a good option.

\noindent$\to$\textit{Ownership model:}
NB-IoT is offered as a connectivity service under a contract that charges a set price for each transmitted byte. The infrastructure is owned by an operator and hence signal coverage depends on the deployed infrastructure, which in turn limits the application owner's control. For example, LoRaWAN allows the user to reduce the energy consumption of the devices by deploying a closer gateway. In addition, applications deployed in remote areas may require other types of networks, such as those enabled by self-managed LoRaWAN gateways.

%\end{itemize}

%-----------------------------------------------------------------------
%-----------------------------------------------------------------------
\ifCLASSOPTIONpeerreview

\else
\begin{comment}
\section*{Acknowledgments}
The research work in this article has been partially supported the SINERGIA project TEC2015-71303-R and the SGR-60-2017 funds.
This project has also received funding from the European Union's Horizon 2020 research 
and innovation programme under grant agreement No.\,726607.
\end{comment}
\fi

\bibliography{nbiot-paper}{}
\bibliographystyle{ieeetr}

\end{document}